\documentclass[10pt,a4paper,onecolumn]{article}
\usepackage{marginnote}
\usepackage{graphicx}
\usepackage[rgb,svgnames]{xcolor}
\usepackage{authblk,etoolbox}
\usepackage{titlesec}
\usepackage{calc}
\usepackage{tikz}
\usepackage[pdfa]{hyperref}
\usepackage{hyperxmp}
\hypersetup{%
    unicode=true,
    pdfapart=3,
    pdfaconformance=B,
    pdftitle={SPARC-X-API: Versatile Python Interface for Real-space
Density Functional Theory Calculations},
    pdfauthor={Tian Tian, Lucas R Timmerman, Shashikant Kumar, Ben
Comer, Andrew J Medford, Phanish Suryanarayana},
    pdfpublication={Journal of Open Source Software},
    pdfpublisher={Open Journals},
    pdfissn={2475-9066},
    pdfpubtype={journal},
    pdfvolumenum={},
    pdfissuenum={},
    pdfdoi={10.xxxxxx/draft},
    pdfcopyright={Copyright (c) 1970, Tian Tian, Lucas R Timmerman,
Shashikant Kumar, Ben Comer, Andrew J Medford, Phanish Suryanarayana},
    pdflicenseurl={http://creativecommons.org/licenses/by/4.0/},
    colorlinks=true,
    linkcolor=[rgb]{0.0, 0.5, 1.0},
    citecolor=Blue,
    urlcolor=[rgb]{0.0, 0.5, 1.0},
    breaklinks=true
}
\immediate\pdfobj stream attr{/N 3} file{sRGB.icc}
\pdfcatalog{%
  /OutputIntents [
    <<
      /Type /OutputIntent
      /S /GTS_PDFA1
      /DestOutputProfile \the\pdflastobj\space 0 R
      /OutputConditionIdentifier (sRGB)
      /Info (sRGB)
    >>
  ]
}
\usepackage{caption}
\usepackage{orcidlink}
\usepackage{tcolorbox}
\usepackage{amssymb,amsmath}
\usepackage{ifxetex,ifluatex}
\usepackage{seqsplit}
\usepackage{xstring}

\usepackage{float}
\let\origfigure\figure
\let\endorigfigure\endfigure
\renewenvironment{figure}[1][2] {
    \expandafter\origfigure\expandafter[H]
} {
    \endorigfigure
}

\usepackage{fixltx2e} 

\NewDocumentCommand\citeproctext{}{}
\NewDocumentCommand\citeproc{mm}{%
  \begingroup\def\citeproctext{#2}\cite{#1}\endgroup}
\makeatletter
 \let\@cite@ofmt\@firstofone
 \def\@biblabel#1{}
 \def\@cite#1#2{{#1\if@tempswa , #2\fi}}
\makeatother
\newlength{\cslhangindent}
\newlength{\csllabelwidth}
\newenvironment{CSLReferences}[2] 
 {\begin{list}{}{%
  \setlength{\itemindent}{0pt}
  \setlength{\leftmargin}{0pt}
  \setlength{\parsep}{0pt}
  \ifodd #1
   \setlength{\leftmargin}{\cslhangindent}
   \setlength{\itemindent}{-1\cslhangindent}
  \fi
  \setlength{\itemsep}{#2\baselineskip}}}
 {\end{list}}
\usepackage{calc}

\usepackage[top=3.5cm, bottom=3cm, right=1.5cm, left=1.0cm,
            headheight=2.2cm, reversemp, includemp, marginparwidth=4.5cm]{geometry}

\titleformat{\section}
  {\normalfont\sffamily\Large\bfseries}
  {}{0pt}{}
\titleformat{\subsection}
  {\normalfont\sffamily\large\bfseries}
  {}{0pt}{}
\titleformat{\subsubsection}
  {\normalfont\sffamily\bfseries}
  {}{0pt}{}
\titleformat*{\paragraph}
  {\sffamily\normalsize}

\usepackage{fancyhdr}
\makeatletter
\let\ps@plain\ps@fancy
\fancyheadoffset[L]{4.5cm}
\fancyfootoffset[L]{4.5cm}

\definecolor{linky}{rgb}{0.0, 0.5, 1.0}

\newtcolorbox{repobox}
   {colback=red, colframe=red!75!black,
     boxrule=0.5pt, arc=2pt, left=6pt, right=6pt, top=3pt, bottom=3pt}

\newcommand{\ExternalLink}{%
   \tikz[x=1.2ex, y=1.2ex, baseline=-0.05ex]{%
       \begin{scope}[x=1ex, y=1ex]
           \clip (-0.1,-0.1)
               --++ (-0, 1.2)
               --++ (0.6, 0)
               --++ (0, -0.6)
               --++ (0.6, 0)
               --++ (0, -1);
           \path[draw,
               line width = 0.5,
               rounded corners=0.5]
               (0,0) rectangle (1,1);
       \end{scope}
       \path[draw, line width = 0.5] (0.5, 0.5)
           -- (1, 1);
       \path[draw, line width = 0.5] (0.6, 1)
           -- (1, 1) -- (1, 0.6);
       }
   }

\definecolor{c53baa1}{RGB}{83,186,161}
\definecolor{c202826}{RGB}{32,40,38}

\patchcmd{\@maketitle}{center}{flushleft}{}{}
\patchcmd{\@maketitle}{center}{flushleft}{}{}
\patchcmd{\@maketitle}{\LARGE}{\LARGE\sffamily}{}{}
\def\maketitle{{%
  
  \AB@maketitle}}
\renewcommand\AB@affilsepx{ \protect\Affilfont}
\renewcommand\AB@affilnote[1]{{\bfseries #1}\hspace{3pt}}
\renewcommand{\affil}[2][]%
   {\newaffiltrue\let\AB@blk@and\AB@pand
      \if\relax#1\relax\def\AB@note{\AB@thenote}\else\def\AB@note{#1}%
        \setcounter{Maxaffil}{0}\fi
        \begingroup
        \let\href=\href@Orig
        \let\protect\@unexpandable@protect
        \def\thanks{\protect\thanks}\def\footnote{\protect\footnote}%
        \@temptokena=\expandafter{\AB@authors}%
        {\def\\{\protect\\\protect\Affilfont}\xdef\AB@temp{#2}}%
         \xdef\AB@authors{\the\@temptokena\AB@las\AB@au@str
         \protect\\[\affilsep]\protect\Affilfont\AB@temp}%
         \gdef\AB@las{}\gdef\AB@au@str{}%
        {\def\\{, \ignorespaces}\xdef\AB@temp{#2}}%
        \@temptokena=\expandafter{\AB@affillist}%
        \xdef\AB@affillist{\the\@temptokena \AB@affilsep
          \AB@affilnote{\AB@note}\protect\Affilfont\AB@temp}%
      \endgroup
       \let\AB@affilsep\AB@affilsepx
}
\makeatother

\renewcommand\Affilfont{\sffamily\small\mdseries}
  \usepackage[T1]{fontenc}
  \usepackage[utf8]{inputenc}

\IfFileExists{upquote.sty}{\usepackage{upquote}}{}
\IfFileExists{microtype.sty}{%
\usepackage{microtype}
\UseMicrotypeSet[protrusion]{basicmath} 
}{}

\PassOptionsToPackage{usenames,dvipsnames}{color} 
\ifLuaTeX
\usepackage[bidi=basic]{babel}
\else
\usepackage[bidi=default]{babel}
\fi
\babelprovide[main,import]{american}

\def\languageshorthands#1{}
\usepackage{color}
\usepackage{fancyvrb}

\DefineVerbatimEnvironment{Highlighting}{Verbatim}{commandchars=\\\{\}}
\newenvironment{Shaded}{}{}

\newcommand{\BuiltInTok}[1]{\textcolor[rgb]{0.00,0.50,0.00}{#1}}

\newcommand{\CommentTok}[1]{\textcolor[rgb]{0.38,0.63,0.69}{\textit{#1}}}

\newcommand{\DecValTok}[1]{\textcolor[rgb]{0.25,0.63,0.44}{#1}}

\newcommand{\FloatTok}[1]{\textcolor[rgb]{0.25,0.63,0.44}{#1}}

\newcommand{\ImportTok}[1]{\textcolor[rgb]{0.00,0.50,0.00}{\textbf{#1}}}

\newcommand{\NormalTok}[1]{#1}
\newcommand{\OperatorTok}[1]{\textcolor[rgb]{0.40,0.40,0.40}{#1}}

\newcommand{\StringTok}[1]{\textcolor[rgb]{0.25,0.44,0.63}{#1}}

\usepackage{draftwatermark}

\usepackage{graphicx,grffile}
\makeatletter
\def\maxwidth{\ifdim\Gin@nat@width>\linewidth\linewidth\else\Gin@nat@width\fi}
\def\maxheight{\ifdim\Gin@nat@height>\textheight\textheight\else\Gin@nat@height\fi}
\makeatother
\setkeys{Gin}{width=\maxwidth,height=\maxheight,keepaspectratio}
\IfFileExists{parskip.sty}{%
\usepackage{parskip}
}{
\setlength{\parindent}{0pt}
\setlength{\parskip}{6pt plus 2pt minus 1pt}
}
\providecommand{\tightlist}{%
  \setlength{\itemsep}{0pt}\setlength{\parskip}{0pt}}
\ifx\paragraph\undefined\else
\let\oldparagraph\paragraph
\renewcommand{\paragraph}[1]{\oldparagraph{#1}\mbox{}}
\fi
\ifx\subparagraph\undefined\else
\let\oldsubparagraph\subparagraph
\renewcommand{\subparagraph}[1]{\oldsubparagraph{#1}\mbox{}}
\fi
\ifLuaTeX
  \usepackage{selnolig}  
\fi

\title{SPARC-X-API: Versatile Python Interface for Real-space Density
Functional Theory Calculations}

\author[1,3%
\ensuremath\mathparagraph]{Tian Tian%
  \,\orcidlink{0000-0003-0634-0455}\,%
}
\author[1%
]{Lucas R Timmerman%
  \,\orcidlink{0000-0001-5664-5762}\,%
}
\author[1%
]{Shashikant Kumar%
  \,\orcidlink{0009-0001-5134-1580}\,%
}
\author[1%
]{Ben Comer%
  \,\orcidlink{0000-0002-7528-0049}\,%
}
\author[1%
\ensuremath\mathparagraph]{Andrew J Medford%
  \,\orcidlink{0000-0001-8311-9581}\,%
}
\author[1,2%
\ensuremath\mathparagraph]{Phanish Suryanarayana%
  \,\orcidlink{0000-0001-5172-0049}\,%
}

\affil[1]{College of Engineering, Georgia Institute of Technology,
Atlanta, GA 30332, USA%
}
\affil[2]{College of Computing, Georgia Institute of Technology,
Atlanta, GA 30332, USA%
}
\affil[3]{Department of Chemical and Materials Engineering, University
of Alberta, Edmonton AB, T6G 2R3, Canada%
}
\affil[$\mathparagraph$]{Corresponding author}
\date{\vspace{-2.5ex}}

\begin{document}
\maketitle

\marginpar{

  \begin{flushleft}
  \sffamily\small

  {\bfseries DOI:} \href{https://doi.org/10.xxxxxx/draft}{\color{linky}{10.xxxxxx/draft}}

  \vspace{2mm}
    {\bfseries Software}
  \begin{itemize}
    \setlength\itemsep{0em}
    \item \href{https://github.com/openjournals}{\color{linky}{Review}} \ExternalLink
    \item \href{https://github.com/openjournals}{\color{linky}{Repository}} \ExternalLink
    \item \href{https://doi.org/10.5281}{\color{linky}{Archive}} \ExternalLink
  \end{itemize}

  \vspace{2mm}
  
    \par\noindent\hrulefill\par

  \vspace{2mm}

  {\bfseries Editor:} \href{https://joss.theoj.org}{Open
Journals} \ExternalLink
   \\
  \vspace{1mm}
    {\bfseries Reviewers:}
  \begin{itemize}
  \setlength\itemsep{0em}
    \item \href{https://github.com/openjournals}{@openjournals}
    \end{itemize}
    \vspace{2mm}
  
    {\bfseries Submitted:} 01 January 1970\\
    {\bfseries Published:} unpublished

  \vspace{2mm}
  {\bfseries License}\\
  Authors of papers retain copyright and release the work under a Creative Commons Attribution 4.0 International License (\href{https://creativecommons.org/licenses/by/4.0/}{\color{linky}{CC BY 4.0}}).

  \end{flushleft}
}

\section{Summary}\label{summary}

Density Functional Theory (DFT) is the de facto workhorse for
large-scale electronic structure calculations in chemistry and materials
science. While plane-wave DFT implementations remain the most widely
used, real-space DFT provides advantages in handling complex boundary
conditions and scaling to very large systems by allowing for the
efficient use of large-scale supercomputers and linear-scaling methods
that circumvent the cubic scaling bottleneck. The SPARC-X project
(\url{https://github.com/SPARC-X}) provides highly efficient and
portable real-space DFT codes for a wide range of first principle
applications, available in both Matlab (M-SPARC
(\citeproc{ref-xu_m-sparc-1.0_2020}{Xu et al., 2020};
\citeproc{ref-zhang_m-sparc-2.0_2023}{Zhang et al., 2023})) and C/C++
(SPARC (\citeproc{ref-xu_sparc-1.0_2021}{Xu et al., 2021};
\citeproc{ref-zhang_sparc-2.0_2024}{Zhang et al., 2024})). The rapid
growth of SPARC’s feature set has created the need for a fully
functional interface to drive SPARC in high-throughput calculations.
Here we introduce SPARC-X-API, a Python package designed to bridge the
SPARC-X project with broader computational frameworks. Built on the
atomic simulation environment (ASE
(\citeproc{ref-larsen_ase_2017}{Hjorth Larsen et al., 2017})) standard,
the SPARC-X-API allows users to handle SPARC file formats and run SPARC
calculations through the same interface as with other ASE-compatible DFT
packages. Beyond standard ASE capabilities, SPARC-X-API provides
additional features including 1) support of SPARC-specific setups,
including complex boundary conditions and unit conversion, 2) a JSON
schema parsed from SPARC’s documentation for parameter validation and
compatibility checks, and 3) a comprehensive socket communication layer
derived from the i-PI protol
(\citeproc{ref-ceriotti_i-pi-1.0_2014}{Ceriotti et al., 2014};
\citeproc{ref-kapil_i-pi-2.0_2019}{Kapil et al., 2019}) facilitating
message passing between low-level C code and the Python interface. The
goal of the SPARC-X-API is to provide a easy-to-use interface for users
with diverse needs and levels of expertise, allowing for minimal effort
in adapting SPARC to existing computational workflows, while also
supporting developers of advanced real-space methods.

\section{Statement of Need}\label{statement-of-need}

DFT has unargubaly become one of the cornerstones of electronic
structure simulations in chemical and materials sciences due to its
simplicity and wide range of applicability. Among the various numerical
implementations of DFT, the plane-wave pseudopotential method has gained
significant popularity, owing to both its robustness and the maturity of
associated software packages. However, despite their widespread use,
plane-wave methods are not without limitations. One long-standing
challenge in DFT is to develop methods that overcomes the huge
computational cost for solving the Kohn-Sham equation, which scales
cubically with respect to the system size. This becomes especially
problematic in massively parallel computing environments, where the
extensive global communication required during Fourier transformations
limits the scalability, making it challenging to efficiently simulate
very large systems in plane-wave DFT. In plane-wave methods, the global
nature of the Fourier basis used limits the ability to achieve linear
scaling (\citeproc{ref-bowler_order_n_dft_2012}{Bowler \& Miyazaki,
2012}). Moreover, the periodic nature of the Fourier basis enforces the
use of periodic boundary conditions, making the simulation setup of
isolated and semi-finite systems non-straightforward. A compelling
alternative to overcome these limitations is to solve the Kohn-Sham
equations using a finite-difference (FD) approach on real-space grids.
The locality of the FD method makes real-space DFT methods inherently
scalable, and paves the way for the development of linearly-scaling
solutions to the Kohn-Sham equations. Real-space DFT also naturally
supports both periodic and Dirichlet boundary conditions, and
combinations thereof, allowing for the flexible treatment of systems in
any dimensionality.

\begin{figure}
\centering
\includegraphics[width=0.9\textwidth,height=\textheight]{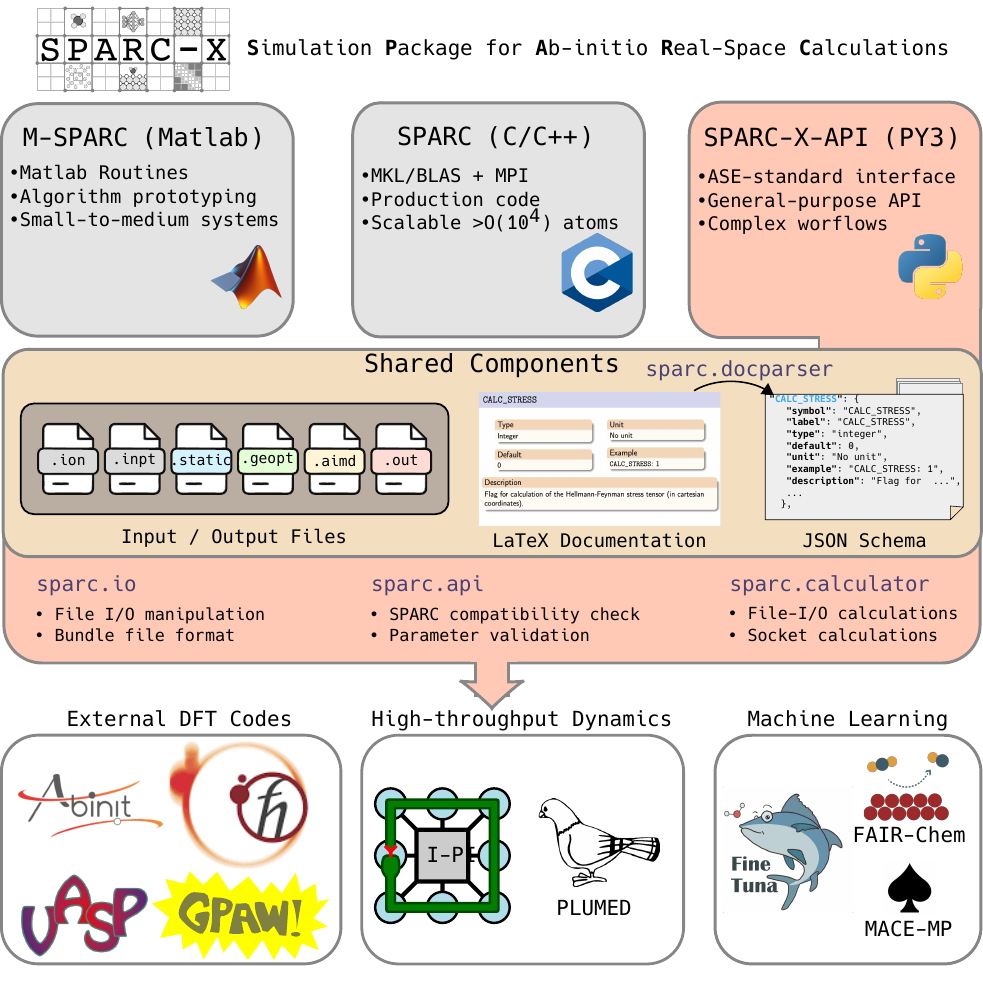}
\caption{Overview of SPARC-X-API in the SPARC-X project system
\label{fig:sparc-overview}}
\end{figure}

In the past few years, the SPARC-X project
(\url{https://github.com/SPARC-X}) has led efforts to develop an
open-source, real-space DFT code that is both user-friendly and
competitive with state-of-the-art plane-wave codes. The philosophy of
the SPARC-X project is to provide codes that are highly efficient and
portable (i.e., straightforward to install and use across various
computational environments). The codes also seek to be user-friendly and
developer-friendly to facilitate the implementation of new algorithms.
In line with this, SPARC-X offers real-space DFT algorithms through two
implementations: 1) Matlab-based M-SPARC
(\citeproc{ref-xu_m-sparc-1.0_2020}{Xu et al., 2020};
\citeproc{ref-zhang_m-sparc-2.0_2023}{Zhang et al., 2023}) for algorithm
prototyping and small-system simulations, with no external dependencies
other than Matlab itself, and 2) C/C++ based SPARC
(\citeproc{ref-xu_sparc-1.0_2021}{Xu et al., 2021};
\citeproc{ref-zhang_sparc-2.0_2024}{Zhang et al., 2024}) for large-scale
production calculations that can accommodate a wide range of system
sizes and requires only MPI and MKL/BLAS for compilation. New features
of SPARC include spin-orbit coupling, dispersion interactions, and
advanced exchange-correlation (xc) functionals
(\citeproc{ref-zhang_sparc-2.0_2024}{Zhang et al., 2024}),
linear-scaling Spectral Quadrature (SQ) method
(\citeproc{ref-suryanarayana_sparc_sq_2018}{Suryanarayana et al.,
2018}), cyclic/helical symmetry
(\citeproc{ref-sharma_sparc_cyclix_2021}{Sharma \& Suryanarayana,
2021}), real-space density functional perturbation theory (DFPT)
(\citeproc{ref-sharma_sparc_dfpt_2023}{Sharma \& Suryanarayana, 2023}),
orbital-free DFT (ODFT) (\citeproc{ref-ghosh_sparc_ofdft_2016}{Ghosh \&
Suryanarayana, 2016}), on-the-fly machine-learning force fields
(OTF-MLFF) (\citeproc{ref-kumar_ofdft_delta_ml_2023}{Kumar et al.,
2023}, \citeproc{ref-kumar_sparc_mlff_2024}{2024};
\citeproc{ref-timmerman_sparc_mlff_2024}{Timmerman et al., 2024}). The
rapid development of SPARC has led to the need for a fully functional
and user-friendly interface that facilitates the use of SPARC with
high-throughput workflows. To address this, we introduce the
SPARC-X-API, a Python interface designed to bridge the SPARC code with a
wide range of scientific workflows. The SPARC-X-API builds upon the
Python wrapper originally shipped with SPARC version 1.0
(\citeproc{ref-xu_sparc-1.0_2021}{Xu et al., 2021}), offering an API
compatible with the widely-used ASE (ASE
(\citeproc{ref-larsen_ase_2017}{Hjorth Larsen et al., 2017})) standard
and updated with the latest versions of SPARC. With ASE’s support for
various popular DFT methods, including both plane-wave (e.g.~VASP
(\citeproc{ref-kresse_vasp_1996}{Kresse \& Furthmüller, 1996}), Quantum
ESPRESSO (\citeproc{ref-giannozzi_qe_2017}{Giannozzi et al., 2017}), and
Abinit (\citeproc{ref-gonze_abinit_2020}{Gonze et al., 2020})), and
real-space (e.g.~GPAW (\citeproc{ref-enkovaara_gpaw_1_2011}{Enkovaara et
al., 2011}; \citeproc{ref-mortensen_gpaw_2_2024}{Mortensen et al.,
2024}) and Octopus
(\citeproc{ref-tancogne_dejean_octopus_2020}{Tancogne-Dejean et al.,
2020})) implementations, SPARC-X-API enables seamless integration of
SPARC into existing workflows, allowing users to incorporate real-space
DFT calculations with minimal adjustments. The modular design of
SPARC-X-API makes it straightforward to be plugged into complex
computational workflows, for example high-throughput dynamics
simulations by i-PI (\citeproc{ref-litman_i-pi-3.0_2024}{Litman et al.,
2024}) and PLUMED (\citeproc{ref-bonomi_plumed_2019}{Bonomi et al.,
2019}), as well as active machine learning frameworks including FineTuna
(\citeproc{ref-musielewicz_finetuna_2022}{Musielewicz et al., 2022}),
powered by state-of-art neural network interatomic potentials such as
FAIR-Chem
(https://github.com/FAIR-Chem/fairchem){[}https://github.com/FAIR-Chem/fairchem{]}
and MACE-MP (\citeproc{ref-ilyes_mace_2023}{Batatia et al., 2024}) model
series. A summary of the role SPARC-X-API in the SPARC-X project is
shown in \autoref{fig:sparc-overview}. In addition to the capabilities
inherited from ASE, SPARC-X-API seeks to enhance the user experience in
a few key aspects, including 1) supporting SPARC-specific features in an
ASE-comatible API, 2) a parameter validation mechanism based on SPARC’s
\texttt{LaTeX} documentation, and 3) a versatile socket communication
layer for efficient high-throughput calculations. Details will be
discussed next.

\section{Features and
Functionalities}\label{features-and-functionalities}

The SPARC-X-API is structured as a Python package, \texttt{sparc}. A
summary of its key functionalities is provided below; for current
detailed documentation, please refer to the
\href{https://sparc-x.github.io/SPARC-X-API}{official documentation}.

\subsection{\texorpdfstring{\texttt{sparc.io}: File I/O
Manupulation}{sparc.io: File I/O Manupulation}}\label{sparc.io-file-io-manupulation}

In SPARC and M-SPARC calculations, input information is provided by two
files: a \texttt{.inpt} (cell dimensions, boundary conditions,
calculation flags), and a \texttt{.ion} file (atomic configurations and
locations to pseudopotential). Depending on the type of calculation,
various output files may be written, such as\texttt{.static},
\texttt{.geopt} or \texttt{.aimd}. The separation of information across
multiple files means converting ASE \texttt{Atoms} objects to SPARC
input files or retrieving energy and forces information from SPARC
calculations requires handling more than just a single file, as is
common in most ASE I/O formats. To manage this, the SPARC-X-API operates
on the directory level, treating each calculation directory as a “SPARC
bundle”. The \texttt{sparc.io.SparcBundle} class facilitates reading
from and writing to this bundle, ensuring that all necessary input and
output files are properly handled. By default, the SPARC-X-API also
copies relevant pseudopotential files into the calculation directory,
making the SPARC bundle portable across different machines. From version
2.0 onwards, the SPARC-X-API leverages the new features introduced in
ASE version 3.23 to register as an external I/O format, allowing reading
and writing SPARC files directly using \texttt{ase.io} submodule:

\begin{Shaded}
\begin{Highlighting}[]
\ImportTok{from}\NormalTok{ ase.io }\ImportTok{import}\NormalTok{ read, write}
\CommentTok{\# 1. Read a SPARC bundle by specifying the \textasciigrave{}sparc\textasciigrave{} format}
\NormalTok{atoms }\OperatorTok{=}\NormalTok{ read(}\StringTok{"sparc\_output\_dir"}\NormalTok{, }\BuiltInTok{format}\OperatorTok{=}\StringTok{"sparc"}\NormalTok{)}
\CommentTok{\# 2. Write to a SPARC bundle from aboth object}
\NormalTok{write(}\StringTok{"sparc\_input\_dir"}\NormalTok{, atoms, }\BuiltInTok{format}\OperatorTok{=}\StringTok{"sparc"}\NormalTok{)}
\end{Highlighting}
\end{Shaded}

The SPARC-X-API also supports parsing complex boundary conditions from
the \texttt{.inpt} file. The periodic (P) and Dirichlet (D) boundary
conditions are translated into \texttt{True} and \texttt{False} values,
respectively, in the corresponding \texttt{pbc} direction of an
\texttt{Atoms} object. Standard ASE objects do not natively support
cyclic (C) or helical (H) boundary conditions that are available in
SPARC, so the SPARC-X-API treats them similarly to Dirichlet boundaries
and stores the original boundary condition information in the
\texttt{info} attribute of the atomic object. This ensures that the
correct boundary combinations are preserved when re-writing to SPARC
input files.

\subsection{\texorpdfstring{\texttt{sparc.api}: Parameter
Validation}{sparc.api: Parameter Validation}}\label{sparc.api-parameter-validation}

In the ASE ecosystem, default calculator interfaces such as
\texttt{FileIOCalculator} do not implement parameter validation, which
can lead to issues such as incorrect parameter settings or
incompatibility when running calculations through ASE. To address this,
the SPARC-X-API introduces a robust parameter validation system using a
JSON schema generated from SPARC’s
\href{https://github.com/SPARC-X/SPARC/tree/master/doc/.LaTeX}{LaTeX
documentation}. A JSON schema contains the version of the SPARC
software, a list of input parameters used in \texttt{.inpt} and
\texttt{.ion} files, as well as supported data types and parameter
categories. Validation is handled via the \texttt{sparc.api.SparcAPI}
class, and includes:

\begin{itemize}
\tightlist
\item
  Verify that the schema is compatible with the version of SPARC binary.
\item
  Convert \texttt{.inpt} fields into Python data types.
\item
  Validate input parameters in both string and numerical formats.
\item
  Output help information about specific parameter(s).
\end{itemize}

Each release of the SPARC-X-API contains a copy of a JSON schema linked
with the latest SPARC release as the default validator, although the
user is can select different combination of SPARC versions and schemas
depending on the version they are using. The separation between the
SPARC-X-API and the core SPARC code not only prevents the need for
hard-coding parameter lists into the API, but also facilitates easier
maintenance: the “central truth” of parameters remains in the SPARC
documentation, maintained by the SPARC core developers, while the
SPARC-X-API focuses on providing a user-friendly interface without being
tied to constant updates. This approach maximizes flexibility and avoids
version conflicts between the API and the underlying code.

\subsection{\texorpdfstring{\texttt{sparc.calculator}:
Socket-Communication Calculator
Interface}{sparc.calculator: Socket-Communication Calculator Interface}}\label{sparc.calculator-socket-communication-calculator-interface}

The submodule \texttt{sparc.calculator} provides a class \texttt{SPARC}
as the main entry point for driving SPARC calculations. This class
provides two modes of operation: 1) a file I/O-based calculator
extending the \texttt{ase.calculators.FileIOCalculator} class, and 2) a
comprehensive socket communication layer that allows direct
communication between the Python API and low-level C/C++ code.

In file I/O mode, the SPARC calculator object utilizes the
\texttt{sparc.io.SparcBundle} for generating input files and
\texttt{sparc.api.SparcAPI} for parameter validation, while the mode of
calculation (single-point, relaxation or molecular dynamics) is
controlled by the input flags. For users transitioning from other DFT
packages and their ASE calculators, the SPARC-X-API is designed to
minimize adaptation effort, but the API is designed to also enable
advanced inputs from expert users. The \texttt{SPARC} calculator class
achieves this by supporting two sets of input parameters: 1) lower-case
special parameters that follow conventions from other ASE DFT
calculators (e.g.~real-space grid spacing \texttt{h} from GPAW, and
exchange-correlation keyword \texttt{xc} from VASP) that use the ASE
default Angstrom-eV system, and 2) case-insensitive raw SPARC input
parameters in Bohr-Hartree units for fine-grained control. This dual
approach is designed so that users familiar with other DFT codes can
adopt SPARC with minimal changes to their existing workflows, while
expert users can exert full control. Basic DFT calculations can be
covered by using standard ASE parameter sets in the SPARC-X-API, as
shown by the side-by-side constructor with VASP and GPAW, using the same
exchange-correlation functional and compatible convergence settings:

\begin{Shaded}
\begin{Highlighting}[]
\CommentTok{\#1. Using VASP}
\ImportTok{from}\NormalTok{ ase.calculators.vasp }\ImportTok{import}\NormalTok{ Vasp}
\NormalTok{calc }\OperatorTok{=}\NormalTok{ Vasp(xc}\OperatorTok{=}\StringTok{"pbe"}\NormalTok{, kpts}\OperatorTok{=}\NormalTok{(}\DecValTok{9}\NormalTok{, }\DecValTok{9}\NormalTok{, }\DecValTok{9}\NormalTok{), ecut}\OperatorTok{=}\DecValTok{450}\NormalTok{, ediff}\OperatorTok{=}\FloatTok{1.e{-}4}\NormalTok{)}

\CommentTok{\#2. Using GPAW}
\ImportTok{from}\NormalTok{ gpaw }\ImportTok{import}\NormalTok{ GPAW}
\NormalTok{calc }\OperatorTok{=}\NormalTok{ GPAW(xc}\OperatorTok{=}\StringTok{"pbe"}\NormalTok{, kpts}\OperatorTok{=}\NormalTok{(}\DecValTok{9}\NormalTok{, }\DecValTok{9}\NormalTok{, }\DecValTok{9}\NormalTok{), h}\OperatorTok{=}\FloatTok{0.25}\NormalTok{, convergence}\OperatorTok{=}\NormalTok{\{}\StringTok{"energy"}\NormalTok{: }\FloatTok{1.e{-}4}\NormalTok{\})}

\CommentTok{\#3. Using SPARC}
\ImportTok{from}\NormalTok{ sparc.calculator }\ImportTok{import}\NormalTok{ SPARC}
\NormalTok{calc }\OperatorTok{=}\NormalTok{ SPARC(xc}\OperatorTok{=}\StringTok{"pbe"}\NormalTok{, kpts}\OperatorTok{=}\NormalTok{(}\DecValTok{9}\NormalTok{, }\DecValTok{9}\NormalTok{, }\DecValTok{9}\NormalTok{), h}\OperatorTok{=}\FloatTok{0.25}\NormalTok{, convergence}\OperatorTok{=}\NormalTok{\{}\StringTok{"energy"}\NormalTok{: }\FloatTok{1.e{-}4}\NormalTok{\})}
\end{Highlighting}
\end{Shaded}

In high-throughput frameworks requiring thousands of single-point DFT
evaluations, relying on file I/O mode can be inefficient, as
calculations are restarted at each DFT call and the total number of
files may exceed SPARC’s default file count limit. The socket layer in
the SPARC-X-API avoids these limitations by directly communicating with
a long-running SPARC process for updating atomic positions, while
keeping density and orbitals in memory and reducing self-consistent
field (SCF) cycles. While alternative communication methods exist, such
as C-binding approaches seen in GPAW
(\citeproc{ref-mortensen_gpaw_2_2024}{Mortensen et al., 2024}) and Psi4
(\citeproc{ref-smith_psi4_2020}{Smith et al., 2020}), these typically
involve complex compilation and integration steps when installing the
Python package. We chose a socket-based communication layer for its
simplicity, which allows for a clear separation between the Python and
SPARC codebases, minimal modifications to the existing C/C++ code, and
ease of installation without requiring recompilation.

\begin{figure}
\centering
\includegraphics[width=1\textwidth,height=\textheight]{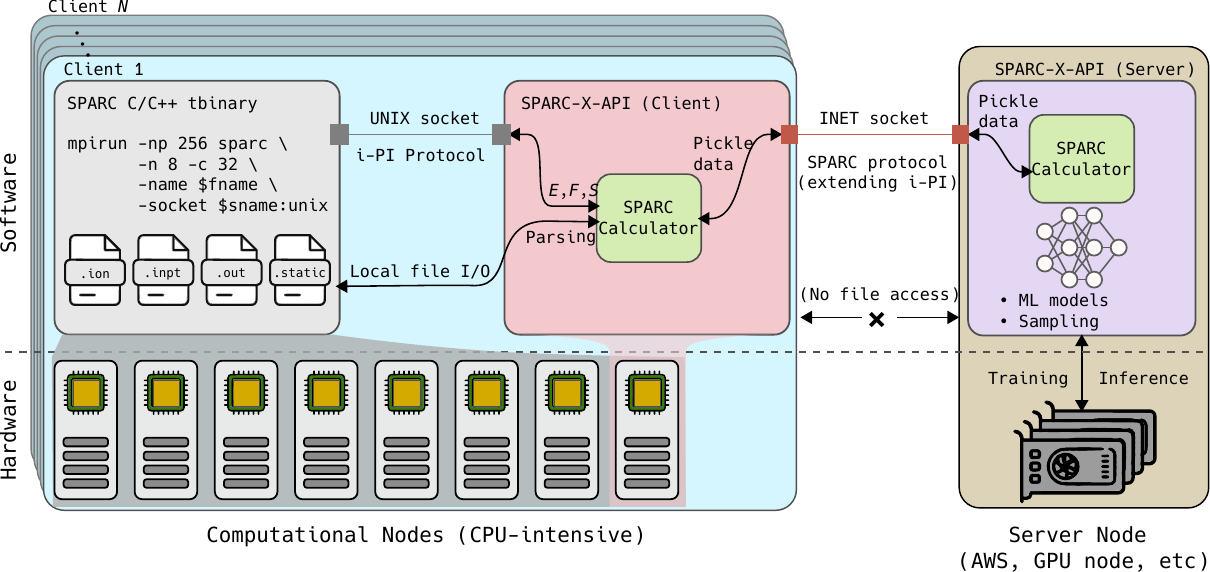}
\caption{Example of socket communication across hybrid computing
platforms using SPARC-X-API \label{fig:socket-hetero}}
\end{figure}

The communication protocol used in the SPARC-X-API socket, referred to
as the SPARC protocol, is based on the i-PI protocol
(\citeproc{ref-ceriotti_i-pi-1.0_2014}{Ceriotti et al., 2014};
\citeproc{ref-kapil_i-pi-2.0_2019}{Kapil et al., 2019}), which is also
adapted by a wide range of ASE calculators. The SPARC protocol
introduces additional header types and supporting binary data transfers
via Python’s pickle format. While SPARC’s C/C++ code maintains
compatibility with the original i-PI standard, the SPARC-X-API leverages
the extended protocol with pickle decoding. The two-tier design offers
flexibility for socket calculations. At its core, the SPARC binary can
communicate directly with any i-PI-compatible server, such as
\texttt{ase.calculators.socketio.SocketIOCalculator} in ASE, using the
basic protocol, though this requires careful setup by the user. However,
the SPARC-X-API leverages the SPARC protocol, which allows the API to
internally relay more advanced data types to the SPARC binary, handling
object decoding and socket resets automatically. When running socket
calculations on a single machine, users can activate socket mode by
simply adding \texttt{use\_socket=True} to the \texttt{SPARC} calculator
constructor, enabling UNIX socket communication without additional
setup. More importantly, the design of the SPARC protocol allows easy
and seamless integration in distributed computational systems, offering
the following features: 1) flexible client initialization / restart 2)
efficient data transfer 3) heterogeneous computational setup. The design
of the SPARC protocol allows insertion of bidirectional additional
routines between two DFT calls, allowing further control over the
low-level C/C++ code. Figure \autoref{fig:socket-hetero} summarizes the
server-client setup across hybrid computing platforms.

\subsection{Miscellaneous Helper
Functionalities}\label{miscellaneous-helper-functionalities}

The SPARC-X-API also provides several helper functions to facilitate
user installation and testing, including:

\begin{itemize}
\tightlist
\item
  \texttt{sparc.quicktest}: a utility to verify the installation and
  environment setups for \texttt{SPARC-X-API} and \texttt{SPARC}.
\item
  \texttt{sparc.docparser}: a submodule to convert existing
  \texttt{LaTeX} documentation included in SPARC source code into JSON
  schema.
\item
  \texttt{sparc.download\_data}: a tool to download the latest ONCV
  pseudopotentials released by SPARC.
\item
  \texttt{sparc-ase}: an extension to the commandline \texttt{ase} tool,
  adding compatibility with SPARC file formats.
\end{itemize}

\section{Code Release and
Maintenance}\label{code-release-and-maintenance}

The SPARC-X-API is released as source code in github repository
\url{https://github.com/SPARC-X/SPARC-X-API}, and as a
\texttt{conda-forge} package
\href{https://anaconda.org/conda-forge/sparc-x-api}{\texttt{sparc-x-api}}.
When installed using \texttt{conda-forge}, the package is bundled with
the optimized SPMS pseudopotentials
(\citeproc{ref-shojaei_sparc_pseudopot_2023}{Shojaei et al., 2023}), and
compatible with the
\href{https://anaconda.org/conda-forge/sparc-x}{\texttt{sparc}} package
that contains the compiled SPARC binary.

It also integrates continuous integration (CI) workflows for:

\begin{itemize}
\tightlist
\item
  Unit testing and code coverage
\item
  Fetching the latest SPARC documentation for updating the JSON schema
\item
  Validating all test examples from the SPARC repository
\end{itemize}

These workflows ensure that SPARC-X-API remains up-to-date with ongoing
SPARC developments while separating parameter updates from the main
SPARC maintainers’ efforts.

\section{Acknowledgements}\label{acknowledgements}

The authors gratefully acknowledge the support of the U.S. Department of
Energy, Office of Science, under Grant No.~DE-SC0019410 and
DE-SC0023445.

\section*{References}\label{references}
\addcontentsline{toc}{section}{References}

\phantomsection\label{refs}

\end{document}